\documentclass{article}
\usepackage{arxiv}
\usepackage{natbib}
\usepackage{graphicx}
\usepackage{multirow}
\usepackage{subcaption}
\usepackage{booktabs}
\usepackage{makecell}
\usepackage{todonotes}
\usepackage{lmodern,url}
\usepackage{siunitx}
\usepackage{xspace}

\title{Peptipedia: a comprehensive database for peptide research supported by Assembled predictive models and Data Mining approaches}
\usepackage{authblk}

\author[1]{Cristofer Quiroz}
\author[2]{Yasna Barrera Saavedra}
\author[3,4]{Benjam\'in Armijo-Galdames}
\author[3,4]{Juan Amado-Hinojosa}
\author[3,4]{\'Alvaro Olivera-Nappa}
\author[3\thanks{\tt{ana.sanchez@ing.uchile.cl}}]{Anamaria Sanchez-Daza}
\author[3,4\thanks{\tt{david.medina@cebib.cl}}]{David Medina-Ortiz}

\affil[1]{Facultad de Ingenier\'ia, Universidad Auton\'oma de Chile, Cinco Pte. 1670, Talca, 3467987, Chile.}
\affil[2]{Escuela de Ingenier\'ia en Bioinform\'atica, Universidad de Talca, Avenida Lircay SN, 3460000, Talca, Chile.}
\affil[3]{Centre for Biotechnology and Bioengineering, University of Chile, Beauchef 851, Santiago, 8370456, Chile.}
\affil[4]{Department of Chemical Engineering, Biotechnology and Materials, University of Chile, Beauchef 851, Santiago, 8370456, Chile.}

\date{}

\renewcommand{\headeright}{ }
\renewcommand{\undertitle}{ }

\begin{document}
\maketitle

\begin{abstract}
\textbf{Motivation}: Peptides have attracted the attention in this century due to their remarkable therapeutic properties. Computational tools are being developed to take advantage of existing information, encapsulating knowledge and making it available in a simple way for general public use. However, these are property-specific redundant data systems, and usually do not display the data in a clear way. In some cases, information download is not even possible. This data needs to be available in a simple form for drug design and other biotechnological applications.\\
\textbf{Results}: We developed Peptipedia, a user-friendly database and web application to search, characterise and analyse peptide sequences. Our tool integrates the information from thirty previously reported databases, making it the largest repository of peptides with recorded activities so far. Besides, we implemented a variety of services to increase our tool's usability. The significant differences of our tools with other existing alternatives becomes a substantial contribution to develop biotechnological and bioengineering applications for peptides.\\
\textbf{Availability}: Peptipedia is available for non-commercial use as an open-access software, licensed under the GNU General Public License, version GPL 3.0. The web platform is publicly available at \url{pesb2.cl/peptipedia}. Both the source code and sample datasets are available in the GitHub repository  \url{https://github.com/CristoferQ/PeptideDatabase}.\\
\textbf{Contact:} david.medina@cebib.cl, ana.sanchez@ing.uchile.cl
\end{abstract}

\keywords{Protein Engineering - predictive models \and peptide databases \and machine-learning algorithms \and digital signal processing \and assembled models}

\section*{INTRODUCTION}

Peptides play a crucial role as signaling molecules, encompassing diverse therapeutic activities like antimicrobial, antitumoral, hormone replacement, anti-inflammatory and antihypertensive \citep{lau2018therapeutic, lien2003therapeutic}. Peptides are polymers that can be sought in natural sources or synthetically obtained; they are constituted of at least 2 amino acids, and their maximum length is usually set to 50 - 100 amino acids. However, it seems there is no consensus about the maximum of amino acids in a sequence to consider it a peptide or a protein. \citep{morrison, latham1999therapeutic, lien2003therapeutic,uhlig2014emergence}. 

As therapeutic agents, peptides are especially attractive because they exhibit high biological activity and specificity, reduced side effects and low toxicity. Nevertheless, peptides have some disadvantages over other molecules, such as high synthesis cost and low stability due of the lack of tertiary structure, making them particularly susceptible to enzymatic degradation and difficulties in crossing biological membranes due to their high polarity, molecular weight, and hydrophilicity.
 \citep{vlieghe2010synthetic,uhlig2014emergence}.

Despite the disadvantages mentioned, peptide researching interest has increased, resulting in a significant accumulation of new peptide sequences in conjunction with their related activities and properties. 
This has brought to the market over 70 peptides approved in the US, Europe, and Japan as therapeutic, more than 200 in clinical trials, and more than 600 in pre-clinical tests \citep{book2019,usmani2017thpdb,lau2018therapeutic}. 

One of the most significant trends in recent times is "drug discovery" to identify new drugs or new functionalities for specific targets. In this context, computational approaches are continually developed as support tools for biological fields, where methodologies based on Machine Learning and Data Mining become relevant tools \citep{wu2019recent,basith2020machine}.  
However, these techniques require prior knowledge, which can be obtained from biological databases that accumulate information on molecules and their characteristics. These data of interest can be collected and processed to develop a tool for solving a specific problem.

Several dedicated databases have emerged for peptide grouping, mostly, according to their activities (e.g., antimicrobial: APD3 \citep{wang2016apd3}, antituberculosis: AntiTBdb \citep{usmani2018antitbpdb} AntiTbPdb, antihypertensive: AHTPDB \citep{kumar2015ahtpdb}) or origin source (e.g., Plant: PlantPepDB \citep{das2020plantpepdb}, bacterial: BACTIBASE \citep{hammami2010bactibase}, anuran: DADP \citep{novkovic2012dadp}). The first web-based databases including peptides were reported in 1998 by \citet{tossi2002molecular}, followed by SYFPEITHI, JenPep, FIMM and HIV database \citep{rammensee1999syfpeithi,blythe2002jenpep,schonbach2000fimm,korber1998hiv}. Then in 2003, the Antimicrobial Peptide Database (APD) appeared and has been continuously updated, but currently, the link is down \citep{wang2004apd, wang2016apd3}, and since then, around 40 peptide databases have arisen.

Each database is useful in their specific context, but a comprehensive and integrated database focused on peptides is not available so far. Also, many of the databases present some issues which hinder their usability. Most of them do not indicate their last update, and if reviewing, they seem to have not been updated since their launch, except for DRAMP, AllergenOnline, BactPepDB, DBAASP, ConoServer and APD. Other sites are not found: PenBase, ANTIMIC.
Almost all databases have redundancy in their sequences (see section 1 of Supplementary Information). Others require informatic background, being unfriendly for users with no advanced computational skills. Many others do not provide a download tool: YADAMP, Quorumpeps database, DADP, BIOPEP, BioDADpep, Péptaibol; for others, the download tool is not working: PepBank, StraPep, PeptideDB, BactPepDB, MHCBN, ForPep, CancerPPD.

Peptipedia was developed to fulfill the necessities that each database cannot solve separately. We have implemented a user-friendly web application with a new database that encompasses the highest number of peptide sequences with reported activity, curated from 30 existing peptide databases. Peptipedia classifies reported activity for each peptide in categories and subcategories defined according to our analysis and literature \citep{kastin2013handbook}. 

Our application is more than a database compilation: it is the most extensive integrated peptide persistent-storage system to date. This user-friendly platform also includes useful physicochemical and statistical properties estimator from peptides, amino acid sequences characterisation, and a tool for Machine Learning-based activity prediction for a query peptide.

\section*{Methods}

\subsection*{Collection, preprocessing,  characterisation, and database generation}

We consolidate the information for Peptipedia by integrating the data from different computational tools and databases previously reported, such as APD \citep{wang2016apd3}, LAMP \citep{zhao2013lamp}, and Uniprot \citep{uniprot2015uniprot}, among others (see section 1 in Supplementary Information for more details). Firstly, we manually downloaded the sequences from each tool and processed them independently, generating different CSV files to facilitate their manipulation. We filtered the sequences according to their length, considering a minimum of 2 residues and a maximum of 150. Secondly, we generated a single file with all sequences, eliminating redundancy between them. For each sequence, we searched its activities, using the previous information in all databases employed to develop our web information system. It is important to note that taxonomic and structural information, and specific information for particular activities, such as IC50 measurements, experiments, among others, were also included in Peptipedia. Furthermore, the sequences are categorized depending on whether they present modifications or non-canonical residues. Then, we used ModLamp library \citep{10.1093/bioinformatics/btx285} to characterise the peptides based on physicochemical and thermodynamic properties. Statistical properties were obtained for each sequence using the DMAKit-Lib library \citep{dmakit}. Finally, the amino acid frequency for each sequence was obtained through scripts implemented in Python v3. Now, we store the processed information in a NoSQL database, using MongoDB as a handler due to its manipulation characteristics, information extraction speed and scaling.

\subsection*{Strategies for classification systems}

Most sequences report a specific activity in terms of their biochemical roles and/or biological effects, specially in humans. We noted that a significant number of peptides are used or were designed for therapeutic purposes, but there were another seven types of peptide activity which cannot be classified as therapeutic. Consequently, we classify all peptides in eight categories: (1) 'therapeutic', (2) 'immunological', (3) 'sensorial', (4) 'neurological', (5) 'drug delivery vehicle', (6) 'transit', (7) 'propeptide' and (8) 'signal'. Each category has subclassifications within it. However, there are a small group of peptides with particular activity, so we categorise them  in the category (9) 'other activity'. All peptides with no activity reported are in the category (10) 'no activity reported'.

One of the essential services of Peptipedia is the activity classification system for peptide sequences based on Machine Learning strategies. The training of models was based on the application of supervised learning algorithms combined with sequence coding approaches, using physicochemical properties and Digital Signal Processing, according to the strategies proposed by \citet{medina2020combination}. In this way, we generated assembled binary models to recognize activities for peptide sequences employing our categories proposed in this work. The training process was based on developing binary data sets to evaluate two categories: presents or absence of activity. Additionally, we generated each data set using the one v/s rest strategy, keeping class imbalance minimum. Finally, in those models with low performance, it was used the recursive binary partitions strategies, according to the method proposed by \citet{medina2020development} to improve the performance of the classification assembled models.

\subsection*{Implementation and Availability}

Peptipedia was designed using a Model View Controller (MVC) design pattern. The view component and the controllers were implemented using JavaScript programming language through the Express framework. Display components were optimised using Bootstrap 4. All the model members, including all service disposed in this work's proposed tool, were developed using Python v3 programming language, supported by the libraries DMAKit-Lib \citep{dmakit} and Scikit-Learn \citep{pedregosa2011scikit}. Both the proposed software architecture and implementation features are detailed in section 2 of the Supplementary Information. 


\section*{Results and Discussion}

\begin{figure*}[!htpb]
    \centering
    \includegraphics[scale=.3]{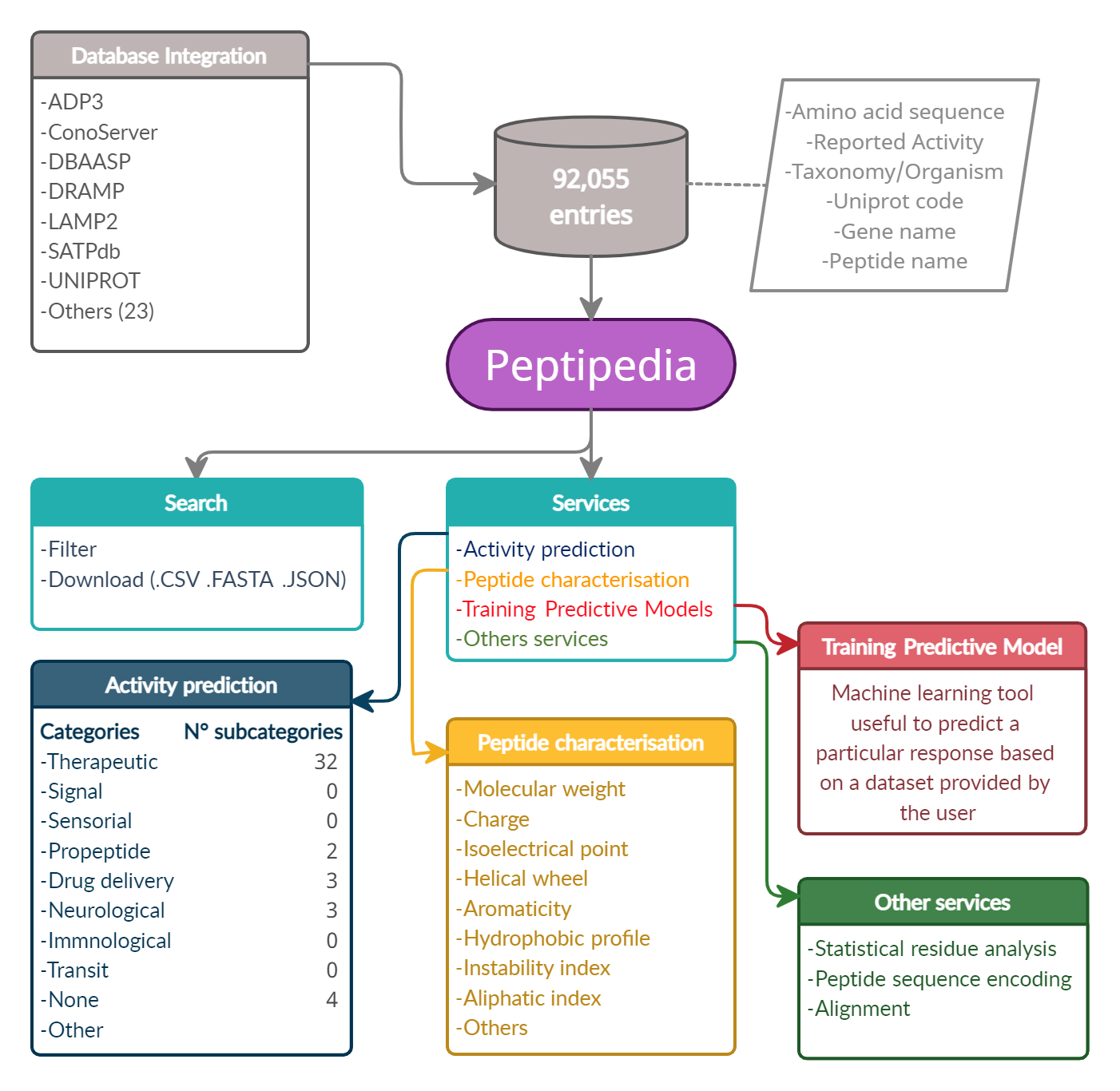}
    \caption{\textbf{Representative scheme of building and characteristics of Peptipedia.} Peptipedia is a computational tool for peptide sequence analysis. The information presented by our tool was consolidated from 30 databases, considering information on the sequence, taxonomy, and different properties of stored peptides. Searching for sequences and relevant information in our web application is easy, personalised and intuitive, allowing download the information in multiple formats. Peptipedia has enabled different tools that will help characterise and analyse sequences, as well as functionalities supported by Machine Learning methods that facilitate the development of predictive models and an activity predictor system.}
    \label{fig:summary}
\end{figure*}

Peptipedia is a user-friendly web application system to search, analyse, evaluate and characterise peptide sequences using different strategies, Machine Learning, and Data Mining techniques. This web tool has a NoSQL database system with 92055 peptides registered and described, being the most extensive database of peptide sequences with activities reported to date. This tool reports different types of information for each sequence, considering structural, physicochemical, and phylogenetic properties. Additionally, the various activities previously identified for each peptide are reported and so are the databases or repositories where they were extracted from. Finally, statistical properties related to the percentage of residues for each sequence and the average per category are included in the database, providing interesting, useful, and easy-to-understand information for scientists and researchers (see Figure \ref{fig:summary}). 

\subsection*{Relevant tools and services available in Peptipedia}

\subsubsection*{Searches,  Visualization and Downloads}

Different types of searches can be generated in Peptipedia, either with the sequence or through information related to its activity, physicochemical properties, frequency of residues, among other relevant information. Besides, it is possible to apply different filters to generate a personalised exploration for the user's interest. 

We develop a general summary for each search, showing statistical descriptions and various visualizations to display the information. Furthermore, we present specific details for each peptide, including thermodynamic properties, taxonomy, phylogeny, activity and sequence descriptors; we also show the databases where the peptide sequence was previously reported. Remarkably, Peptipedia offers specific information like IC50, assays information, organism evaluation and other relevant characteristics for particular activities such as antihypertensive, anti-HIV, and antiviral subcategories. 

Peptipedia has general and specific modules for downloading data, making it easier to obtain information, facilitating the download in CSV, Fasta, and JSON formats. Besides, our tool enables the complete database download in easily manipulable forms, considering both the sequence and its reported information.

\subsubsection*{Services}

Different services were implemented in Peptipedia to facilitate analysis and characterisation of peptide sequences. We propose various services that allow characterization through physicochemical and thermodynamic properties, using the ModLamp \citep{10.1093/bioinformatics/btx285} library. We also provide modules that enable the estimation of statistical properties for peptide sequences. 

Bioinformatic tools such as sequence alignments are available in our web tool: using the Edlib library \cite{10.1093/bioinformatics/btw753}, it is possible to align any sequence against those registered in our database. 

Another relevant service is peptide activity classification system supported by assembled predictive models: the user can upload a list of amino acid sequences, and our tool classifies them by the categories proposed in this work, evaluating each of them. Furthermore, a peptide encoding service is implemented using common strategies such as One Hot Encoder and more sophisticated ones such as Embedding through the Tape library. 

Finally, Peptipedia allows the generation of predictive models for sets of peptides with specific user requirements through supervised learning algorithms and cross-validation techniques. Configuration of hyperparameters, coding strategy  and  validation method are selectable. The tool reports the performance of the generated model by the user, allowing the download to use it locally. Besides, this service enables the interpretation of the results giving different recommendations about them. 

\subsection*{Peptides registered, categories, and relevant information in Peptipedia}

\begin{figure*}[!htpb]
    \centering
    \includegraphics[width=16cm, height=9cm]{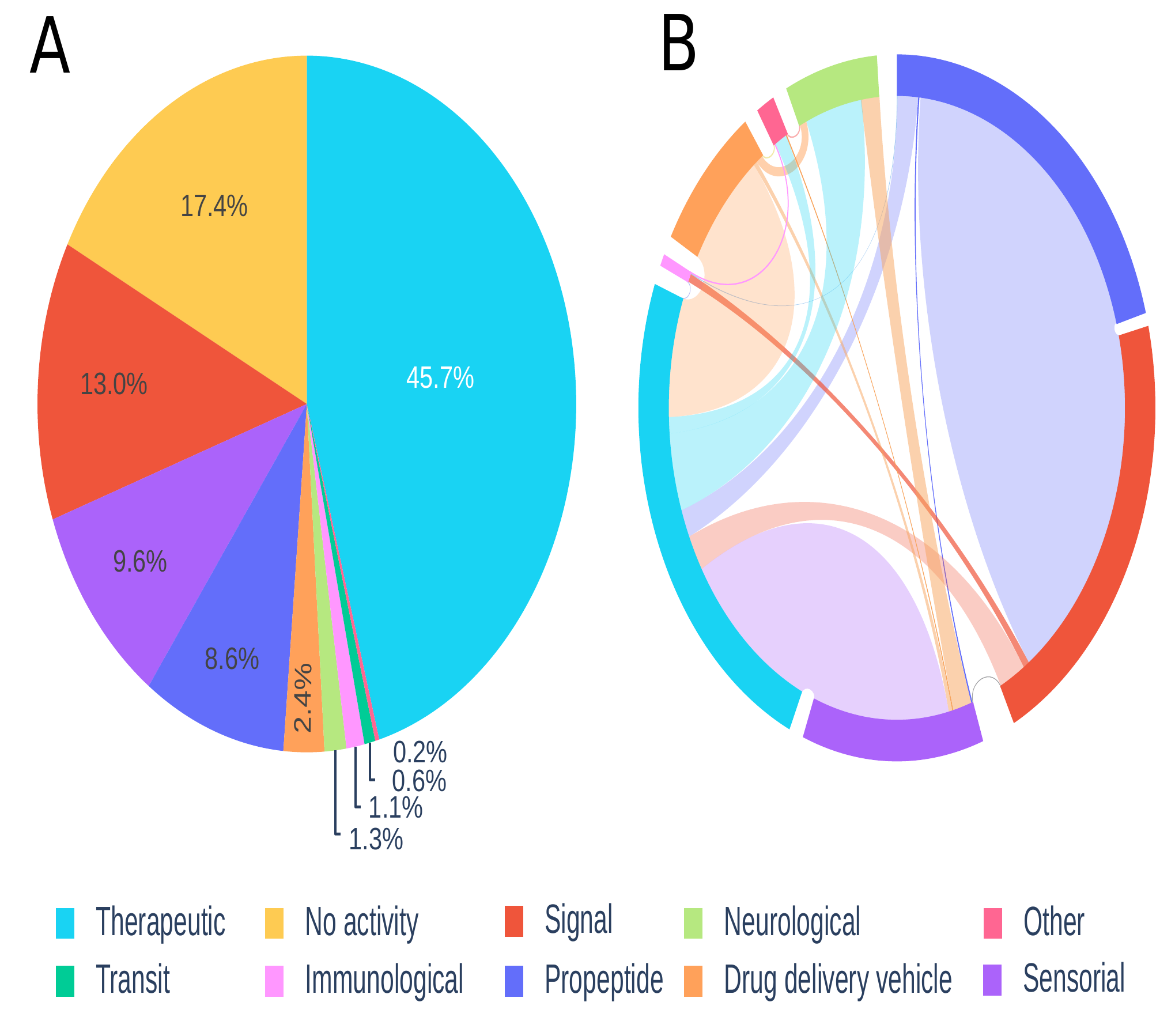}
    \caption{\textbf{Visualisation of registered peptides on Peptipedia} Representation of the information contained in Peptipedia. A: distribution of peptides according to the categories proposed in this work. B: analysis of the relationship of simultaneous activities for the same type of peptide; the most significant trends are seen between therapeutics and sensorial, and between propeptides and signal.}
    \label{fig:fig01}
\end{figure*}

We developed the largest database of peptides with reported activity to date, with a total of 92,055 records. Considering the information on previously reported activities and the characteristics of each their specific properties, we propose a system of ten categories, which present sub-categories according to the features of the activities that constitute them. Using these categories, we analyse the peptide sequences, identifying therapeutic peptides, signal peptides, and sensory activity, representing the highest prevalence in our records. While immunological, transit, and neurological activity show the least trend or have fewer records (see Figure \ref{fig:fig01} A).

It is important to highlight the moonlighting characteristics of peptides. This feature is the feasibility of a peptide to present different activities at the same time \citep{jeffery1999moonlighting}. The main found tendencies of moonlighting are between the therapeutic and sensorial peptides, and between propeptides and  signal peptides. This last overlapping of activities makes sense because propeptides generally contain a signal peptide in their sequence \citep{wang_strapep_2018}, which they lose once processed. (see Figure \ref{fig:fig01} B). This type of properties reflects the potential features of a peptide when acting as a drug or presenting different biotechnological uses, making them interesting to study due to their fascinating characteristics. Residue frequency analysis allows evaluating amino acid trends for particular activities. We compare trends for the main reported categories, with a clear preference for arginine residues for drug delivery peptides, which can be explained because this kind of peptides are usually design to crosss membranes, so they need a chemical affinity for negatively charged membranes, which is given by the positive charge of arginine. In contrast, signal, transit and propeptides generally show similar trends. However, no major visible patterns were identified (see Section 4 in Supplementary Information).

\subsection*{Binary classification categories supported by Assembled Models}

Using coding of physicochemical properties and their representation in frequency space \citep{medina2020combination} and employing recursive binary division strategies to optimise performance measures \citep{medina2020development} we depeloped 44 assembled binary models for classification of activity for peptide sequences, considering the categories and subcategories proposed in this work. We used k-fold cross-validation to avoid model overfitting. Remarkably, all the models generated presented an accuracy of over 83\% (see Table \ref{tab:tab_performance} and section 5 of the Support Information for details). We previously compared the results obtained by applying this type of strategies against classical sequence coding methods, demonstrating better results \citep{medina2020combination}. Furthermore, we compare our results with previously developed classification models for peptide sequences. \citet{xiao2013iamp} proposed a classification system for antimicrobial peptides with 86\% accuracy; for the same task, our model achieves a performance of 88.7\%. Similarly, \citet{yi2019acp} proposed a classification system for anticancer peptides using Deep Learning Long Short-Term Memory Model strategies, achieving an accuracy of 81.48\%, while our model achieves 83.54\%. Another relevant example is identifying quorum sensing peptides (QSPs): \citet{rajput2015prediction} proposed an identification system for QSPs based on sequence features in combination with support vector machine algorithms, obtaining 93 \% accuracy; our accuracy is slightly lower for this peptides, reaching an accuracy of 86.4 \%.  Even though we present a lower performance in particular situations than previously developed methods. Nevertheless, the proposed strategy is generic, could be apply in activity classification of peptides sequences problems, prediction of properties, and multiple issues in protein engineering \citep{medina2020combination}. Notably, we validated all our models using statistical methods. Each data set was created by selecting random samples and repeating this process 100 times, providing statistical support and demonstrating the robustness of the activity classification models implemented in Peptipedia.

\begin{table}[!htpb]
\centering
\begin{tabular}{llll}
\hline
\multicolumn{1}{c}{\textbf{\#}} &
  \multicolumn{1}{c}{\textbf{Category}} &
  \multicolumn{1}{c}{\textbf{\begin{tabular}[c]{@{}c@{}}Size\\ dataset\end{tabular}}} &
  \multicolumn{1}{c}{\textbf{\begin{tabular}[c]{@{}c@{}}Weighted \\ Performance\end{tabular}}} \\ \hline
1. & Sensorial Peptides & 19982 & 85.27 \\
2. & Drug Delivery      & 4912  & 86.02 \\
3. & Therapeutic        & 50000 & 87.32 \\
4. & Neurological       & 2712  & 89.33 \\
5. & Immunological      & 2178  & 86.12 \\
6. & Other Activity     & 490   & 82.98 \\
7. & Transit Peptide    & 1350  & 88.48 \\
8. & Signal Peptide     & 26794 & 86.41 \\
9. & Propeptide         & 17768 & 88.63 \\ \hline
\end{tabular}%

\caption{Weighted performance for binary classification models for the nine main categories proposed in this work.}
\label{tab:tab_performance}
\end{table}

\subsection*{Using Peptipedia to develop predictive models}

\begin{figure}[!htpb]
    \centering
    \includegraphics[width=8cm]{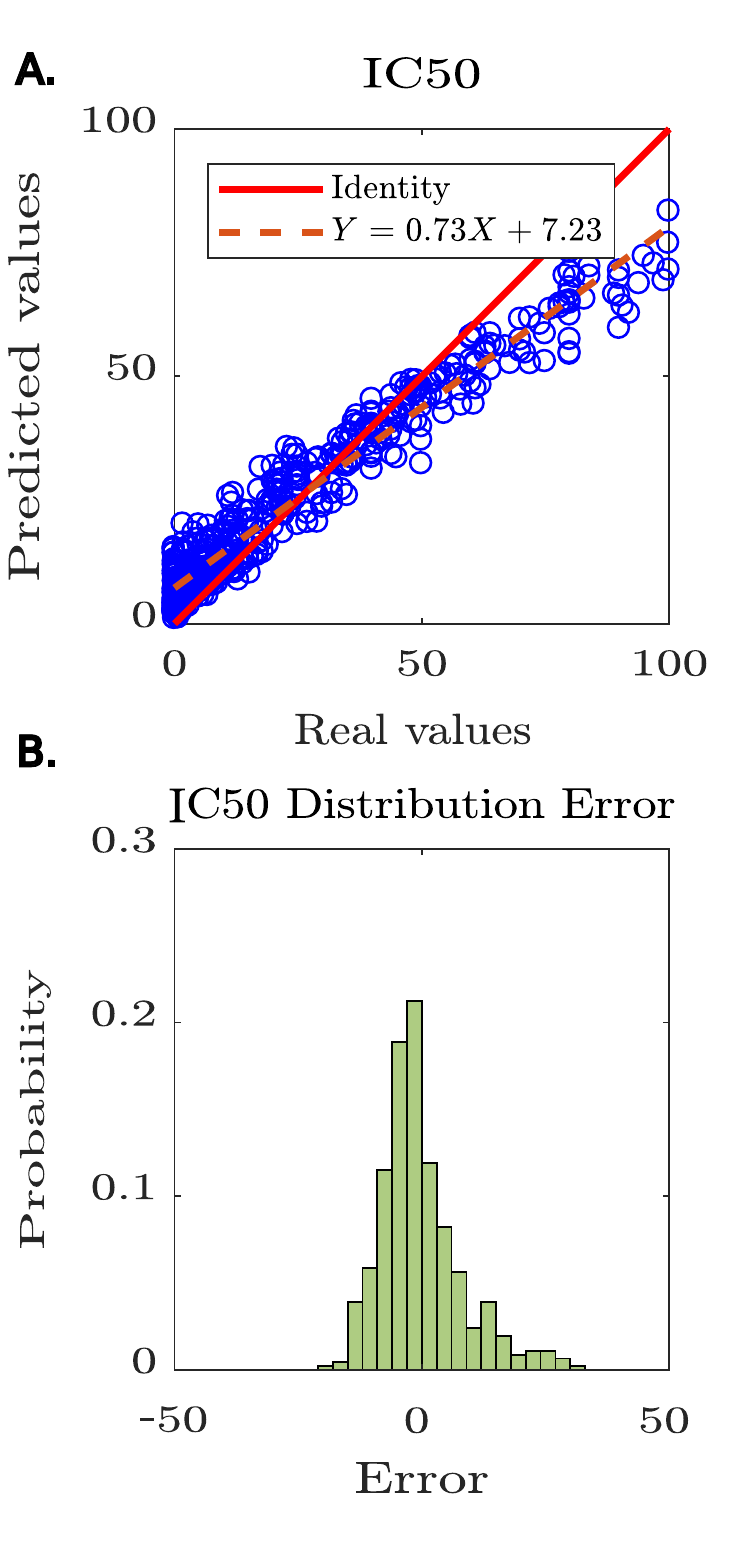}
    \caption{\textbf{Predictive modeling of IC50 for Anti-HIV peptides using Peptipedia} A: Scatter Plot prediction v/s reality, denoting the performance of the predictive model. In general, there is no tendency to over-adjust or under-adjust in any particular range, which shows that the cross-validation strategies were correctly applied. B: histogram of the error distribution. The probability of error analysis indicates no tendency for significant errors that adversely alter the model predictions. The errors are mainly concentrated between -5 and 5, which is quite acceptable considering the nature of the entered values, where the largest reach 100 and the smallest are close to zero.}
    \label{fig:case_study}
\end{figure}

The study of anti-HIV peptides is relevant because their potential therapeutic applications. They interact with a specific domain of the glycoprotein 41, which is their pharmacological target for inhibiting the virus fusion and entry to the host cell. Different efforts have focused on designing new sequences, either through traditional techniques such as directed evolution or rational design strategies. Both strategies currently benefit from the application of Machine Learning since it facilitates the simulation of the effects of new variants. We implemented an IC50 predictive model for anti-HIV peptides to demonstrate the usability of Peptipedia, because this is a crucial parameter for assessing the performance of antimicrobial and antiviral drugs. First, using the sequence search engine, we identify all the peptides that have this category. We manually downloaded and filtered those with a quantitative IC50 measurement, discarding the cases in which it was expressed in terms of low, medium or high effect, and standardising the measured values to work with them using the same units. Subsequently, we used the Peptipedia predictive models training tool, selecting coding by digital signal processing, using the alpha-structure property as coding strategy, Random Forest as supervised learning algorithm, and validation strategy $k$-fold with $k=10$. The tool reported the model's performance, achieving a Pearson coefficient of 0.8 (see Figure \ref{fig:case_study} A). Furthermore, Peptipedia allows us to analyse the prediction error's randomness to determine if there are biases in the generated predictions (see Figure \ref{fig:case_study} B). In this way, we are able to predict the therapeutic potency of an new anti-HIV peptide with no need of performing lab assays, which, combined with the coding module, becomes powerful support for designing peptides with desirable activities.

\section*{Conclusions}

We designed and implemented Peptipedia, a web application supported by machine learning algorithms and data mining strategies  to characterise and analyse peptide sequences. Additionally, our tool has the most extensive database of peptides with activity reported so far, with a total of 92,055 amino acid sequences integrated from thirty databases or repositories of previously reported peptides, Peptipedia has enabled different tools that will help characterising, getting statistical properties and bioinformatics analysis supported by sequence alignments, as well as services that facilitate the development of predictive models.

Additionally, the sequence and the reported activity information of the registered peptides are integrated into a robust binary classification system, implemented through Machine Learning strategies, allowing to predict putative peptide activities. These services are useful as a previous approach to experimental work for performing an activity screening of novel peptides with unknown activity. Besides, peptide design also gets benefited, since this tool helps to find residues patterns based on their activity.

Both the usability and the wide range of services available on Peptipedia, as well as the robustness of the predictive systems implemented, considerably improve the current state of the art, becoming an attractive alternative to existing traditional applications and a good support for research in peptide engineering and its biotechnological applications.

\section*{CODE AVAILABILITY}

All code is available at the authors' GitHub repository \url{https://github.com/CristoferQ/PeptideDatabase}.

\section*{ACKNOWLEDGEMENTS}

This work was supported mainly by the Centre for Biotechnology and Bioengineering - CeBiB (PIA project FB0001, ANID, Chile), Fondecyt 1180882 project, and Universidad de Magallanes for MAG1895 project. DM-O gratefully acknowledges ANID, Chile, for Ph.D. fellowship 21181435. JA-H gratefully acknowledges ANID, for Ph.D. fellowship 21182109. AS-D thanks PAI Programme (I7818010006).

\subsubsection*{Conflict of interest statement.} None declared.

\bibliographystyle{apalike}
\bibliography{document_bib}

\end{document}